\documentclass[11pt]{article}
\usepackage[margin=1in]{geometry}
\usepackage{amsmath,amssymb}
\usepackage{graphicx}
\usepackage{booktabs}
\usepackage{hyperref}
\usepackage{listings}
\usepackage{xcolor}
\usepackage{cite}
\usepackage{microtype}
\usepackage{caption}
\usepackage{float}

\hypersetup{
  colorlinks=true,
  linkcolor=blue,
  citecolor=blue,
  urlcolor=blue
}

\newcommand{\repourl}{https://github.com/Binoculars-X/neuro-fabric}
\newcommand{\bfsixteenloss}{0.020}
\newcommand{\gpuvaloss}{1.5224}
\newcommand{\cpubfvaloss}{1.5426}
\newcommand{\repourltext}{github.com/Binoculars-X/neuro-fabric}
\newcommand{\repoversion}{v1.1.0}
\newcommand{\repocommit}{e9ab47a}

\title{\textbf{NeuronFabric: A Software Reference Architecture\\
for On-Chip Transformer Training with Local Adam}\\[0.5em]
\large BF16W Weights, Vocabulary Budget, and a Path to FPGA Training without a Host CPU}

\author{
  Evgeny Ukladchikov \\
  Independent Researcher \\
  \texttt{ev.uklad@procoders.com.au} \\[0.3em]
  \small \href{\repourl}{\repourltext}
}

\date{June 2026}

\begin{document}

\maketitle

% ─────────────────────────────────────────────────────────────────────────────
\begin{abstract}
% ─────────────────────────────────────────────────────────────────────────────

To our knowledge, publicly documented accelerator architectures generally separate
training compute from optimizer state updates or rely on external memory/host
orchestration. Inference-only chips (Groq LPU, Apple ANE, IBM NorthPole) do not
support on-chip gradient computation; training accelerators (Cerebras, Tenstorrent)
perform the weight update off-chip.
Publicly documented training systems generally separate gradient computation from optimizer-state storage or orchestration.

This paper describes an attempt to fix that. We built a complete C\# software prototype
of a transformer that runs forward pass, backpropagation, and Adam weight update
entirely in one process with no external framework. The purpose is to establish that
the math works and the numbers are right, before committing to silicon.

The model we trained is a 334K parameter autoregressive transformer
($d{=}88$, $H{=}4$ heads, $f{=}264$, $L{=}4$ layers, vocab$=256$) trained on the
Shakespeare corpus. The BF16W variant reaches eval loss \cpubfvaloss{} within 80K samples
(GPU FP32 oracle: \gpuvaloss{}), with coherent character-level text generation confirmed.

We also introduce \textbf{BF16W}: storing weights in BF16 while keeping Adam moments
in FP32. For the target FPGA budget, BF16W becomes practically necessary to leave SRAM headroom for activations.
A 334K FP32 model with Adam moments requires 4.0 MB --- exactly the ZCU102 BRAM limit,
leaving zero headroom for activation buffers. With BF16W it requires 3.34 MB,
leaving 660\,KB free.

We describe the vocabulary-budget constraint we discovered during earlier experiments,
quantify the BF16W SRAM saving, and lay out the FPGA training target as the next step.
No FPGA measurements are included in this paper; FPGA implementation and measurement are left for future work.

\textbf{Code:} \href{\repourl}{\repourltext} (release \repoversion, commit \texttt{\repocommit}).

\textbf{Publication purpose:} This paper serves as a public architectural disclosure
and software reference implementation for future FPGA/ASIC implementations of the
NeuronFabric system, establishing prior art for the local Adam update architecture.
NeuronFabric is a research prototype and not a production accelerator.

\end{abstract}

% ─────────────────────────────────────────────────────────────────────────────
\section{Introduction}
% ─────────────────────────────────────────────────────────────────────────────

Training large language models at the scale of GPT-4 has been estimated at tens of megawatts
sustained over months~\cite{patterson2022carbon}.
A central architectural reason for this is \emph{where the weight update happens}.

Existing neural accelerators separate the compute from the update: gradients are shipped
off-chip and the optimizer runs on a host CPU or GPU.
The optimizer update is typically orchestrated separately from the compute unit that produced the activations.

The gap this creates is concrete:

\begin{itemize}
  \item \textbf{Weight traffic.} GPU inference loads all weights from HBM on every
    forward pass. For a 70B model that is $\sim$140\,GB of memory traffic per sample.
  \item \textbf{Training architecture.} To our knowledge, publicly documented training
    accelerators perform the weight update off-chip or on a host processor.
    The chip that computed the gradients does not apply them.
  \item \textbf{No continuous learning.} An inference chip cannot update from new data
    without sending everything to an external GPU cluster and reloading --- a cycle
    that takes minutes to hours even for fine-tuning.
\end{itemize}

NeuronFabric is an attempt to close that gap. The long-term goal is a silicon chip
where every neuron holds its own weights and runs Adam locally --- weights update in
place, minimizing off-chip optimizer-state and gradient traffic, scaling achieved by connecting chips via
activation links only. This paper presents a software reference implementation demonstrating convergence under the tested configuration. FPGA implementation of the same training loop is the immediate next step.

\subsection{Contributions}

\begin{enumerate}
  \item A complete C\# implementation of a Pre-LN transformer with full backpropagation
    and local Adam, with no external ML framework dependency.
    Training, evaluation, and checkpointing run in a single binary.

  \item \textbf{BF16W} --- a mixed-precision scheme where weights are stored as BF16
    (2 bytes) while Adam moments $m$ and $v$ remain FP32 (4 bytes each). We show this
    reduces SRAM from 12 bytes/parameter to 10 bytes/parameter with a negligible
    convergence penalty ($+\bfsixteenloss$ val loss) on Shakespeare.

  \item The \textbf{vocabulary-budget constraint}: at small parameter budgets, the
    embedding table dominates the model and the transformer layers have almost nothing
    left to learn with. We quantify this across three domains and give the design
    implication for fixed-budget hardware.

  \item A concrete FPGA training target for future work: the 334K BF16W Shakespeare model
    (vocab$=256$) fits entirely in ZCU102 on-chip SRAM and can train without touching DDR.
\end{enumerate}

\subsection{Related Work}

\textbf{Inference-only ASICs.}
Groq LPU~\cite{groq}, Apple ANE~\cite{apple_ane}, and IBM NorthPole~\cite{ibm_northpole}
achieve excellent inference efficiency through SRAM-local weight storage and dataflow
architectures. None support on-chip gradient computation.

\textbf{Training accelerators.}
Cerebras WSE-3~\cite{cerebras} and Tenstorrent Wormhole~\cite{tenstorrent} support
training but the weight update still happens off-chip --- gradients are aggregated
on a host CPU or external optimizer.

\textbf{Neuromorphic chips.}
Intel Loihi 2~\cite{loihi2} and SpiNNaker~\cite{spinnaker} support local weight updates
via spike-timing-dependent plasticity (STDP) --- a Hebbian rule that does not implement
gradient descent. STDP-based learning systems have not demonstrated transformer-scale
gradient training comparable to backpropagation-based LLM training.

% ─────────────────────────────────────────────────────────────────────────────
\section{Architecture}
% ─────────────────────────────────────────────────────────────────────────────

The software architecture is structured so that every abstraction has a direct
hardware analogue. The hierarchy is:

\[
\texttt{NeuronCore} \;\subset\; \texttt{AttentionCore} \;\subset\;
\texttt{AttentionLayer} \;\subset\; \texttt{TransformerBus}
\]

\subsection{NeuronCore: Weight Storage + Compute Unit}

A \texttt{NeuronCore} holds its own weight vector $\mathbf{w} \in \mathbb{R}^n$
and bias $b$. On the forward pass it computes a dot product; on the backward pass
it writes an updated weight in place. There is no weight bus between cores --- each
core owns its weights permanently.

\textbf{Forward:}
\begin{equation}
z = \mathbf{w}^\top \mathbf{x} + b, \qquad y = \sigma(z)
\end{equation}

\textbf{Backward and weight update (Adam):}
\begin{align}
\delta &= g \cdot \sigma'(z) \\
m &\leftarrow \beta_1 m + (1-\beta_1)\,\delta\,\mathbf{x} \\
v &\leftarrow \beta_2 v + (1-\beta_2)\,(\delta\,\mathbf{x})^2 \\
\hat{m} &= m / (1-\beta_1^t), \quad \hat{v} = v / (1-\beta_2^t) \\
\mathbf{w} &\leftarrow \mathbf{w} - \eta \cdot \hat{m} / (\sqrt{\hat{v}} + \epsilon)
\end{align}

The weight update is a direct in-place write. On FPGA this maps to a BRAM write
on the \texttt{Backward} control signal --- no off-chip traffic.

\textbf{BF16W variant:} $\mathbf{w}$ is stored as \texttt{ushort} (BF16), cast to
FP32 for computation, and cast back after the Adam step. Moments $m$ and $v$ remain FP32.
This is the configuration used for all experiments in this paper.

\subsection{Transformer Architecture (334K Shakespeare Config)}

\begin{table}[ht]
\centering
\begin{tabular}{lr}
\toprule
Hyperparameter & Value \\
\midrule
Embedding dim $d$ & 88 \\
Attention heads $H$ & 4 \\
Head dim $d_h = d/H$ & 22 \\
Feedforward dim $f$ & 264 ($= 3d$) \\
Transformer layers $L$ & 4 \\
Sequence length $T$ & 128 \\
Vocabulary & 256 (byte-level) \\
\midrule
Total parameters & 334K \\
\bottomrule
\end{tabular}
\caption{334K Shakespeare model — canonical configuration}
\label{tab:arch}
\end{table}

Each transformer layer uses Pre-LN residual connections~\cite{xiong2020layer}:

\begin{align}
x' &= x + \text{MultiHeadAttn}\bigl(\text{LN}_1(x)\bigr) \\
x'' &= x' + \text{FF}\bigl(\text{LN}_2(x')\bigr)
\end{align}

The feedforward block uses GeLU activation:
$\text{FF}(x) = W_2 \cdot \text{GeLU}(W_1 x)$, where $W_1 \in \mathbb{R}^{d \times f}$
and $W_2 \in \mathbb{R}^{f \times d}$.

\textbf{Weight tying:} The output projection reuses the embedding matrix transposed:
$\text{logits}[t, v] = \text{layerOut}[t] \cdot E[v]$.
This removes $|V| \times d = 256 \times 88 = 22{,}528$ parameters from the output head
with no accuracy cost --- an important saving at small budgets.

\subsection{Parameter Count Breakdown}

\begin{table}[ht]
\centering
\begin{tabular}{lrr}
\toprule
Component & Formula & Params \\
\midrule
Embedding & $|V| \times d = 256 \times 88$ & 22,528 \\
Attention per layer & $4 \times d \times d_h \times H = 4 \times 88 \times 22 \times 4$ & 30,976 \\
Feedforward per layer & $2 \times d \times f = 2 \times 88 \times 264$ & 46,464 \\
LayerNorm (approx) & $\sim$200 per layer & 800 \\
\midrule
Per layer total & & 77,440 \\
$\times$ 4 layers & & 309,760 \\
Embedding (weight-tied) & & 22,528 \\
\midrule
\textbf{Total} & & \textbf{$\sim$334K} \\
\bottomrule
\end{tabular}
\caption{Parameter breakdown for 334K Shakespeare model}
\label{tab:param-breakdown}
\end{table}

\subsection{Hardware Mapping}

The design intent is that every software abstraction maps directly to a hardware block:

\begin{table}[H]
\centering
\begin{tabular}{ll}
\toprule
Software & Hardware target \\
\midrule
\texttt{NeuronCore} weight array & BRAM block \\
\texttt{NeuronCore.Forward()} & DSP48 multiply-accumulate chain \\
\texttt{NeuronCore.Backward()} & BRAM write (same block, no bus) \\
\texttt{NeuronLayer} parallel dispatch & Parallel \texttt{always} blocks in Verilog \\
Softmax \texttt{exp()} & 256-entry LUT in BRAM\textsuperscript{$\dagger$} \\
\texttt{TransformerBus} control enum & 3-bit on-chip control signal\textsuperscript{$\ddagger$} \\
\bottomrule
\end{tabular}
\caption{Software-to-hardware mapping (FPGA target).
\textsuperscript{$\dagger$}256 entries cover the stable-softmax argument range (attention logits after max-subtraction are bounded, typically $[-30, 0]$); range decomposition or interpolation may be required for full BF16 accuracy --- deferred to FPGA implementation.
\textsuperscript{$\ddagger$}Control signal is internal to the chip; host interface design (SPI, AXI-Lite, or similar) is deferred to FPGA implementation.}
\end{table}

The control bus uses six states: \texttt{Idle / Encode / Forward / Backward /
WeightRead / WeightWrite} --- a 3-bit signal that drives all compute blocks.
This is not an abstraction convenience; it is the intended silicon interface.

\begin{figure}[H]
\centering
\includegraphics[width=\linewidth]{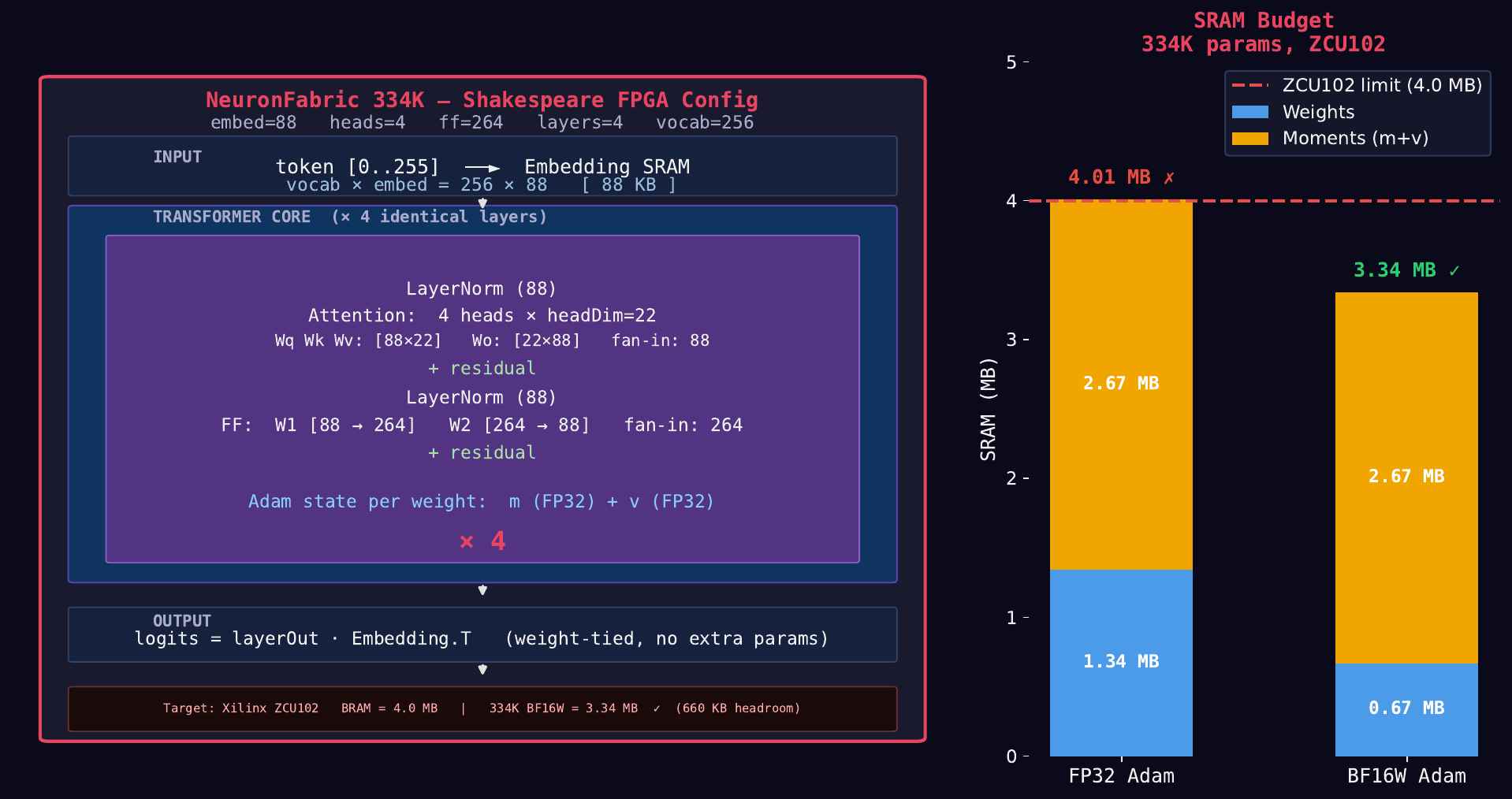}
\caption{NeuronFabric 334K Shakespeare FPGA configuration (embed=88, heads=4, ff=264,
layers=4, vocab=256). \textbf{Left}: chip architecture — token embedding SRAM (88\,KB),
four Pre-LN transformer layers each with local Adam state, weight-tied output projection.
\textbf{Right}: SRAM budget on ZCU102 (4.0\,MB BRAM). FP32 Adam requires 4.00\,MB
(no headroom for activations); BF16W Adam requires 3.34\,MB, leaving 660\,KB free.
Weights: FP32 = 1.34\,MB, BF16W = 0.67\,MB. Moments ($m$+$v$, FP32 in both) = 2.67\,MB.}
\label{fig:arch-334k}
\end{figure}

% ─────────────────────────────────────────────────────────────────────────────
\section{BF16W: Why Weight Compression Is Not Optional for FPGA}
% ─────────────────────────────────────────────────────────────────────────────

\label{sec:bf16w}

The FPGA training target requires all weights \emph{and} Adam moments to live in
on-chip SRAM during training. The Xilinx Zynq UltraScale+ ZCU102 has 32.1\,Mb
($\approx 4.0$\,MB) of BRAM. The 334K model in FP32 with Adam moments needs:

\[
334{,}000 \times (4 + 4 + 4) \text{ bytes} = 334{,}000 \times 12 = 4.00\,\text{MB}
\]

That barely fits. Adding activation buffers for backpropagation ($\sim 180$\,KB per layer,
recomputed layer-by-layer) pushes the FP32 total over the limit.

\textbf{BF16W drops weight storage from 4 bytes to 2 bytes:}

\[
334{,}000 \times (2 + 4 + 4) \text{ bytes} = 334{,}000 \times 10 = 3.34\,\text{MB}
\]

This gives 660\,KB of headroom for activation buffers, control state, and the
embedding SRAM. The model fits comfortably.

\begin{table}[H]
\centering
\begin{tabular}{lrrr}
\toprule
Precision & Weights & Moments ($m$+$v$) & Total \\
\midrule
FP32 Adam & 1.34 MB & 2.67 MB & 4.00 MB \\
\textbf{BF16W Adam} & \textbf{0.67 MB} & \textbf{2.67 MB} & \textbf{3.34 MB} \\
\midrule
ZCU102 BRAM & & & 4.00 MB \\
\bottomrule
\end{tabular}
\caption{SRAM budget for 334K model. FP32 Adam fills the ZCU102 BRAM exactly with no
headroom for activations. BF16W Adam leaves 660\,KB free --- sufficient for activation
buffers and control state.}
\end{table}

\textbf{BF16W implementation.} Weights are stored as \texttt{ushort} arrays.
Each training step: cast BF16 $\to$ FP32, compute forward + backward in FP32,
apply Adam update in FP32, round FP32 $\to$ BF16, write back.
Moments $m$ and $v$ stay in FP32 throughout --- this is where precision matters most
for Adam stability. The weight values themselves tolerate BF16 range ($\sim$3 decimal
digits) because the Adam step size is small relative to weight magnitude.

\textbf{Convergence.} On the 80K-sample Shakespeare runs, CPU BF16W
reaches eval loss \cpubfvaloss{} versus GPU FP32 \gpuvaloss{} --- a gap of $+\bfsixteenloss$.
Both variants converge qualitatively identically; the residual gap is consistent
with the difference in compute precision and random seed (Section~\ref{sec:experiments}).

% ─────────────────────────────────────────────────────────────────────────────
\section{The Vocabulary-Budget Constraint}
% ─────────────────────────────────────────────────────────────────────────────

\label{sec:vocab}

During earlier experiments with smaller models (100K parameters), our results were
consistently worse than expected --- not because backpropagation was wrong (the
gradient checks passed), but because the embedding table was consuming most of the
parameter budget before the transformer layers had any say.

\textbf{The constraint.} For a model with total parameters $P$, embedding dimension
$d$, and vocabulary size $|V|$, the remaining capacity for transformer reasoning is:

\begin{equation}
P_{\text{reason}} = P - |V| \cdot d
\label{eq:vocab-budget}
\end{equation}

We call the term $|V| \cdot d$ the \textbf{vocabulary tax}. With weight tying, the
output projection costs nothing extra --- but the embedding itself is unavoidable.

\textbf{Empirical evidence} from our earlier 100K param experiments:

\begin{table}[H]
\centering
\begin{tabular}{lrrrr}
\toprule
Domain & $|V|$ & Vocab tax & $P_\text{reason}$ & Final loss \\
\midrule
Appointment (GPT-4 synth.) & 49   & 3.1K  & 96.9K & \textbf{0.42} \\
MultiWOZ 2.2               & 302  & 19.3K & 80.7K & 2.05 \\
TinyStories                & 1501 & 96.1K & 3.9K  & 2.90 \\
\bottomrule
\end{tabular}
\caption{Vocabulary-budget constraint across three domains at 100K parameter budget
($d{=}64$). When $P_\text{reason} < 20$K the model produces recognisable words in
incoherent order.}
\label{tab:vocab-budget}
\end{table}

The TinyStories result is the clearest illustration: the model has 1501 vocabulary
tokens but only 3.9K parameters in the transformer layers. It recognises English
words but cannot form coherent sentences. Once $P_{\text{reason}}$ crosses $\sim$80K
the model generates recognisable structural patterns; at 97K it produces fluent
domain text.

\textbf{Implication for the 334K model.} At vocab$=256$ and $d{=}88$:
\[
P_{\text{reason}} = 334{,}000 - 256 \times 88 = 334{,}000 - 22{,}528 = 311{,}472
\]
The vocabulary tax is only 6.7\% of the parameter budget. This is why byte-level
Shakespeare is a better fit than word-level datasets at this scale: the transformer
layers have 311K parameters to work with, not 4K.

\textbf{Hardware design implication.} A fixed-budget FPGA chip must either
specialise to a small-vocabulary domain or separate the embedding onto a dedicated
chip (shared-embedding MoE, described briefly in Section~\ref{sec:future}).

% ─────────────────────────────────────────────────────────────────────────────
\section{Experiments}
% ─────────────────────────────────────────────────────────────────────────────

\label{sec:experiments}

All experiments were conducted using NeuronFabric release \textbf{\repoversion}.
All runs were performed on Windows 11 with an AMD Ryzen 9 9900X CPU and NVIDIA GeForce RTX 4090 GPU (CUDA 13.2, .NET 10.0).

\subsection{Correctness Validation}

Before any corpus training we validated backpropagation using numerical gradient
checks: computing $\partial L / \partial w$ analytically (backward pass) and via
finite differences across all components --- \texttt{NeuronCore},
\texttt{AttentionCore}, \texttt{AttentionLayer}, and \texttt{TransformerBus}.
All implemented gradient checks pass within configured numerical tolerance. More than a hundred unit and regression tests additionally cover the Adam update rule, BF16W moment
accumulation, and the attention softmax backward path independently.
The gradient check is the test that cannot be passed by tuning: it computes the
same derivative two different ways, and if backpropagation is wrong it fails
regardless of how reasonable the output looks.

\subsection{Shakespeare Char-Level Training (334K, Adam)}

\textbf{Dataset.} The Shakespeare corpus: 1,039,854 training characters,
115,540 validation characters. Byte-level tokenisation, vocab$=256$.

\textbf{Configuration.} Architecture as in Table~\ref{tab:arch}.
Optimizer: Adam with linear LR schedule, warmup 200 steps, peak LR $= 0.003$.
Training: 80,000 samples (online, batch$=1$).
Two variants were run: \textbf{GPU Adam FP32} (NVIDIA RTX 4090, CUDA 13.2) as the oracle
baseline, and \textbf{CPU Adam BF16W} (AMD Ryzen 9 9900X, no GPU required) to validate
the mixed-precision scheme.
Checkpoint format: \texttt{.neuro} (JSON header + flat binary weights), version-stamped.

\textbf{Train/validation split.} The Shakespeare corpus is split 90/10:
1,039,854 characters for training, 115,540 for validation.
The model never sees the validation set during training.
All loss and accuracy numbers reported here are measured on the held-out validation set,
not the training set --- the goal is generalisation, not memorisation.

\begin{table}[H]
\centering
\begin{tabular}{lrrrrrr}
\toprule
Variant & Samples & Val loss & BPC & Accuracy & ms/sample \\
\midrule
GPU Adam FP32  & 80,000 & \textbf{\gpuvaloss} & 2.20 & 55.66\% & 17.13 \\
    CPU Adam BF16W & 80,000 & \cpubfvaloss & 2.23 & 54.91\% & 139.04 \\
\midrule
Karpathy char-rnn~\cite{karpathy2015charrnn} & --- & --- & $\sim$1.3 & --- & --- \\
\bottomrule
\end{tabular}
\caption{Shakespeare 334K training results (NeuronFabric \repoversion, run/gpu-fp32-shakespeare-334k-b1-80k/ and run/cpu-bf16w-shakespeare-334k-b1-80k/).
GPU Adam FP32 is the oracle baseline.
CPU BF16W gap: $+\bfsixteenloss$ val loss, $8.1\times$ slower --- consistent with the precision difference between BF16 and FP32 weight storage.
BPC = bits per character = val loss $/ \ln 2$. Karpathy char-rnn reference is a $\sim$3M parameter LSTM; the gap is expected at our 9$\times$ smaller budget and confirms convergence rather than SOTA quality.}
\label{tab:shakespeare-results}
\end{table}

\begin{figure}[H]
\centering
\includegraphics[width=0.85\linewidth]{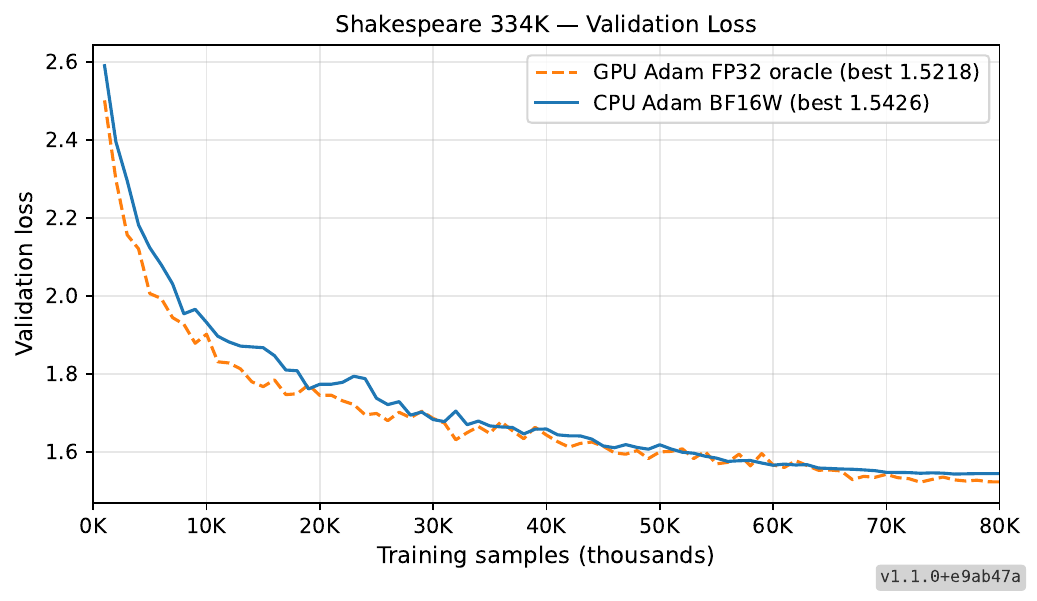}
\caption{Validation loss vs training samples for Shakespeare 334K model.
GPU Adam FP32 reaches eval loss \gpuvaloss{} at 80K samples (oracle baseline).
CPU Adam BF16W reaches \cpubfvaloss.}
\label{fig:shakespeare-loss}
\end{figure}

\textbf{Training curve.} Best eval loss \gpuvaloss{} (GPU FP32, at 80K) and \cpubfvaloss{} (CPU BF16W, at 76K)
were reached within the 80K sample budget. Neither variant overfit --- eval and train loss
tracked closely throughout. Figure~\ref{fig:shakespeare-loss} shows both curves.

\textbf{Sample output} (CPU Adam BF16W, temperature 0.8, 80K samples):
\begin{lstlisting}
> HAMLET:

HAMLET:
Break him of him; for that will for straight,
The both speak ancible o'er an to him all in.

Third Citizens:
Here it field at be that were more cames,
And then is only back'd lawly fortune our peint.
Indlow. For my name more.

DUKE VINCENTIO:
It my straight souls, but mine so rebost 's be
\end{lstlisting}

The model produces recognizable Shakespeare-style dialogue structure and named-character
patterns. This is not state-of-the-art quality: our 334K parameter model reaches
\textbf{2.23 BPC} (bits per character, = eval loss / $\ln 2$ = $\cpubfvaloss / 0.6931$),
compared to Karpathy's char-rnn~\cite{karpathy2015charrnn} at $\sim$1.3 BPC with
$\sim$3M parameters. The gap is expected given our 9$\times$ smaller model and is
not our claim --- the claim is that the implemented local Adam training configuration converges on the Shakespeare benchmark.

% ─────────────────────────────────────────────────────────────────────────────
\section{Hardware Feasibility and Next Steps}
% ─────────────────────────────────────────────────────────────────────────────

\label{sec:future}

\subsection{FPGA Training Target}

The 334K BF16W Shakespeare model is our concrete FPGA training target. No FPGA
measurements are included in this paper --- we are reporting it here to make the
claim falsifiable and to explain why BF16W is necessary rather than optional.

The key properties that make this model FPGA-friendly:

\begin{enumerate}
  \item \textbf{Fits in BRAM.} 3.34 MB weights + moments $< 4.0$ MB ZCU102 BRAM.
    No DDR traffic during training.
  \item \textbf{Batch$=1$ online training.} Adam updates one sample at a time.
    Activation memory scales with one layer at a time (recomputed during backward),
    not with batch size. Activation recomputation adds compute but no extra SRAM.
  \item \textbf{No host CPU in the update loop.} The weight update completes on the
    same SRAM that holds the weights. The host sends a token sequence and receives
    a loss value. Everything in between happens on chip.
\end{enumerate}

The target result for the FPGA implementation: loss drops from $\sim$3.2 (random init) to $\sim$1.54
entirely on the FPGA, matching the software result in Table~\ref{tab:shakespeare-results},
with no host CPU involvement in the Adam update step.

\textbf{DSP throughput analysis (under stated assumptions).}
At a target clock domain of 150--200\,MHz and assuming non-blocking local SRAM access,
the ZCU102 DSP48E2 blocks can sustain a high BF16 FMA utilization rate for the
dot-product chains in each attention head and feedforward layer.
Local SRAM eliminates the memory-bandwidth bottleneck that would otherwise serialize
DSP utilization in off-chip-memory designs.
This is an architectural throughput analysis under stated assumptions, not a measured result.
No FPGA timing closure has yet been demonstrated.
Actual timing and power will be reported after FPGA implementation.

\subsection{Scaling Beyond One Chip}

For larger vocabularies or deeper models the vocabulary-budget constraint
(Section~\ref{sec:vocab}) motivates a multi-chip approach: a dedicated embedding chip
pays the vocabulary tax once, and expert chips contain only transformer layers.
Inter-chip traffic is the hidden state vector $\mathbf{h} \in \mathbb{R}^{T \times d}$
only --- fixed regardless of model depth or total parameter count.

\begin{equation}
\text{bytes per sample} = T \times d \times 4 = 128 \times 88 \times 4 = 45{,}056 \text{ bytes}
\end{equation}

This is not a GPU-style gradient bus. Gradients do not cross chip boundaries.
Each chip updates its own weights locally. Scaling is achieved by adding chips,
not by widening a communication bus.

\subsection{Silicon Roadmap}

The BF16W scheme is process-independent: 10 bytes per trainable parameter regardless of node.
As on-chip SRAM density improves with process scaling, the same architecture accommodates
larger models without architectural change. Silicon-level projections are left for future work
pending FPGA validation.

A possible long-term architecture is a network of chips exchanging activations rather
than optimizer state: each chip is an autonomous learning unit, connected via activation
links only, with no gradient bus and no host CPU in the update loop.

% ─────────────────────────────────────────────────────────────────────────────
\section{Discussion}
% ─────────────────────────────────────────────────────────────────────────────

\subsection{What This Paper Does and Does Not Claim}

To be direct: this paper does not have FPGA measurements. The claim that NeuronFabric
can train a transformer entirely on chip rests on software experiments and FPGA
architectural analysis, not on a running FPGA system. FPGA validation is future work.

What this paper does establish:

\begin{itemize}
  \item The backpropagation implementation is numerically validated via gradient checks
    (finite differences vs.\ analytical gradient across all components), backed by more
    than a hundred unit and regression tests.
  \item BF16W Adam converges with a negligible gap ($+\bfsixteenloss$ val loss) versus FP32 Adam on Shakespeare at 80K samples.
  \item The 334K BF16W model fits in ZCU102 BRAM with 660\,KB headroom --- this is
    arithmetic, not a projection.
  \item The vocabulary-budget constraint (Section~\ref{sec:vocab}) is observed consistently across three exploratory domains and is supported by the arithmetic of equation~\ref{eq:vocab-budget}.
\end{itemize}

\subsection{Reproducibility}

All results in this paper are reproducible from the published codebase.
Every architectural claim is backed by a test that either passes or fails.
The gradient check in Section~\ref{sec:experiments} is the hardest to fake ---
it computes the same derivative two different ways. Training runs are logged
to \texttt{.log} files alongside each \texttt{.neuro} checkpoint, with full
hyperparameters and per-sample loss recorded.

To reproduce the Shakespeare 334K results from scratch:
\begin{enumerate}
  \item \texttt{run/build.bat} --- builds all targets (auto-detects CUDA)
  \item \texttt{cd run/gpu-fp32-shakespeare-334k-b1-80k} then \texttt{1.train.bat} --- GPU Adam FP32 oracle
  \item \texttt{cd run/cpu-bf16w-shakespeare-334k-b1-80k} then \texttt{1.train.bat} --- CPU BF16W (no CUDA required)
  \item \texttt{2.demochat.bat} --- interactive generation from checkpoint
  \item \texttt{3.plot.bat} --- loss curve (PNG + PDF)
\end{enumerate}

Each experiment folder is self-contained: \texttt{\_{}config.bat} holds all
hyperparameters; the numbered scripts read from it. Checkpoint and log are
saved inside the experiment folder (\texttt{checkpoint.neuro},
\texttt{checkpoint.neuro.log}).

The vocabulary-budget experiments in Table~\ref{tab:vocab-budget} (Section~\ref{sec:vocab})
are illustrative results from earlier exploratory runs and are not fully reproduced by the
published scripts. The datasets used (GPT-4 synthetic appointments, MultiWOZ 2.2,
TinyStories) and the 100K-parameter configuration are described in the paper
but the corresponding training scripts are not included in the \repoversion{} release.
The Shakespeare 334K results in Table~\ref{tab:shakespeare-results} are fully reproducible
from the published codebase.

% ─────────────────────────────────────────────────────────────────────────────
\section*{Acknowledgements}

A preprint of this paper will be deposited on arXiv; a Zenodo/DOI archive will be
created at that time to provide a citable, timestamped snapshot of the software and results.

% ─────────────────────────────────────────────────────────────────────────────
\section{Conclusion}
% ─────────────────────────────────────────────────────────────────────────────

We built a complete software reference implementation of a transformer that trains
entirely on a single machine, with local Adam, in pure C\#, with no external ML
framework. We trained a 334K BF16W model on Shakespeare and got coherent text at
eval loss \cpubfvaloss{} --- with 660\,KB of on-chip SRAM headroom to spare on a mid-range FPGA.

The next step is to run the same training loop on the FPGA and get the same loss curve.
If that works, the claim that on-chip transformer training is viable --- without a host
CPU in the weight update loop --- becomes a measured fact rather than a software projection.

% ─────────────────────────────────────────────────────────────────────────────
% Inline references (to be moved to refs.bib before submission)

\end{document}